\documentclass[12pt,english]{amsart}
\usepackage[T1]{fontenc}
\usepackage[latin9]{inputenc}
\usepackage{verbatim}
\usepackage{textcomp}
\usepackage{amstext}
\usepackage{amsthm}
\usepackage{amssymb}

\makeatletter
\numberwithin{equation}{section}
\numberwithin{figure}{section}
\theoremstyle{plain}
\newtheorem{thm}{\protect\theoremname}
  \theoremstyle{definition}
  \newtheorem{example}[thm]{\protect\examplename}

\makeatother

\usepackage{babel}
  \providecommand{\examplename}{Example}
\providecommand{\theoremname}{Theorem}

\begin{document}

\title{Examples of reflection positive Euclidean field theories}

\author{R. Trinchero}
\begin{abstract}
The requirement of reflection positivity(RP) for Euclidean field theories
is considered. This is done for the cases of a scalar field, a higher
derivative scalar field theory and the scalar field theory defined
on a non-integer dimensional space(NIDS). It is shown that regarding
RP, the analytical structure of the corresponding Schwinger functions
plays an important role. For the higher derivative scalar field theory
RP does not hold. However for the scalar field theory on a NIDS, RP
holds in a certain range of dimensions where the corresponding Minkowskian field is defined on a Hilbert
space with a positive definite scalar product that provides a unitary
representation of the Poincaré group. In addition, and motivated by
the last example, it is shown that, under certain conditions, one
can construct non-local reflection positive Euclidean field theories
starting from the corrected two point functions of interacting local
field theories.
\end{abstract}

\date{14/07/2017}
\maketitle

\section{Introduction}

Given a Euclidean field theory defined by its $n$-point Schwinger
functions, the Osterwalder-Schröder (OS) axioms\cite{Osterwalder1973}
allow to reconstruct a field theory on Minkowski space satisfying Wightman
axioms. One of the OS axioms involves the condition of reflection
positivity(RP), this condition enables the definition of a positive
semi-definite scalar product in Minkowski space. This, together with
the other axioms, make possible to produce a set of Wightman functions which transform under unitary representations of the Poincaré group.
 In this work standard and non-standard examples of two point Schwinger
functions are considered. These are the free scalar field, a higher
derivative regulated scalar field and a scalar field in a NIDS. The
example involving the higher derivative scalar field shows that the
pole structure of the two point function plays a important role in
determining the states of this system and their norms. The idea of extending
the number of dimensions to non-integers   appears in physics
in two different fields, the Wilson renormalization group\cite{Wilson:1973jj} and dimensional 
regularization\cite{bollini1972dimensional}\cite{tHooft:1972fi}. From the mathematical point of view a general theory that
can describe NIDS is  given by non-commutative geometry\cite{connes1995noncommutative}. The application of 
non-commutative geometry to  NIDS's as required for the applications\footnote{For its application   to fractal
systems see \cite{connes1995noncommutative},\cite{lapidus-08}} mentioned above 
appears in\cite{Trinchero:2012zn}.  In that reference 
a self-interacting scalar field is considered, the dimension $d$
of that NIDS is given by,
\begin{equation}
d=\frac{n}{1-2\alpha}\label{eq:da}
\end{equation}
where $n$ is the integer dimension of the space that is deformed
and $\alpha\in\mathbb{R}$ the parameter controlling the deformation.
In that reference it is shown that when $\alpha\neq0$ the one-loop
diagrams in these spaces are finite and the singularity structure
for $\alpha\to0$ is the same as in dimensional regularization. An
important point, which was not addressed in that reference, is whether
the theory in that space is unitary or not. In other words whether
one can define a physical theory in these spaces or that they only
provide a regularization of the corresponding integer dimensional
theories. As mentioned above reflection positivity is a key property
in this respect and that is the reason for considering its validity.
It turns out that RP holds in this case for $0<\alpha<\frac{1}{n+2}$.
For $\alpha<0$ the space does not fulfill the conditions required by NCG,
however one can consider the resulting field theory for these $\alpha$ values.
It is shown that for $\alpha<0$ RP does not hold.

In addition, unlike the integer dimensional case the two point function
in the NIDS in momentum space presents a branch cut in the complex
$p^{2}$ plane. The appearance of this branch cut is similar to what
happens in a interacting integer dimensional local field theory when
considering loop corrections to the two point function. Motivated
by this similarity the following question is addressed: Under what
conditions the corrected Euclidean propagator coming from a local
interacting field theory leads to a scalar product satisfying the
requirement of reflection positivity?

The features and results of this work are summarized as follows, 
\begin{itemize}
\item Examples of Euclidean field theories are considered and the condition
of RP is checked for each case.
\item For the higher derivative scalar field, which involves more than one
pole for the two point function in momentum space, the results makes
sense if each pole is considered separately.
\item RP does not hold for the case of the scalar field in the NIDS if $\alpha<0$.
However  RP holds  for $0<\alpha<\frac{1}{n+2}$ . For this case the two point function 
in momentum space presents a branch cut, similar to what happens in a interacting local field theory.
\item Motivated by the similarity mentioned in the previous item, it is
shown that a corrected propagator in a local interacting unitary field
theory leads a scalar product satisfying RP whenever the particle
associated to that field can not decay.
\end{itemize}
The paper is organized as follows. Section \ref{sec:Reflection-positivity}
presents some general facts about RP and checks this condition for
each of the above mentioned cases. Section \ref{sec:Scalar-product-and}
presents the scalar product and energy for each case. Section \ref{sec:New-reflection-positive}
studies the conditions under which a perturbatively defined interacting
and unitary local field theory leads to a corrected propagator satisfying
RP. Section \ref{sec:Concluding-remarks} collects some concluding
remarks.

\section{Reflection positivity\label{sec:Reflection-positivity} }

In the framework of Euclidean functional integrals\cite{glimm1987quantum},
reflection positivity can be phrased as follows,
\begin{equation}
<(\Theta\mathcal{F}[\varphi])\mathcal{F}[\varphi]>\geq0\label{eq:rp}
\end{equation}
where $\mathcal{F}[\varphi]$ denotes a functional of the fields with
support in a region whose points have $x_{d+1}>0$ and $\Theta\mathcal{F}[\varphi]$
denotes the same functional complex conjugated and reflected respect
to the $x_{d+1}=0$ plane. The brackets denote the expectation value
respect to $S[\varphi]$, i.e.,
\[
<(\Theta\mathcal{F}[\varphi])\mathcal{F}[\varphi]>=\frac{1}{\int\mathcal{D\varphi}\,e^{-S[\varphi]}}\int\mathcal{D\varphi}\,e^{-S[\varphi]}(\Theta\mathcal{F}[\varphi])\mathcal{F}[\varphi]
\]
For a free theory or a perturbatively defined interacting theory,
it suffices to consider the case when the functional $\mathcal{F}[\varphi]$
is linear\footnote{Reflection positivity for higher order monomials follows using Wick's
theorem.} in $\varphi$ and $S[\varphi]$ is quadratic in the field and its
derivatives. This linear functional is written as,
\[
\mathcal{F}[\varphi]=\int d^{d+1}x\,f(x)\varphi(x)
\]
where $f(x)$ has support on points $x\in\mathbb{R}_{d+1}^{+}$, where
$\mathbb{R}_{d+1}^{+}$ denotes the points of $\mathbb{R}_{d+1}$
such that $x_{d+1}>0$. Defining the functional,
\[
G[J]=\int\mathcal{D\varphi}\,e^{-S_{J}[\varphi]}\;\;,\;S_{J}[\varphi]=S[\varphi]+\int d^{d+1}x\,J\varphi
\]
the expectation value appearing in the l.h.s. of (\ref{eq:rp}) is,
\begin{align*}
<(\Theta\mathcal{F}[\varphi])\mathcal{F}[\varphi]> & =\left.\frac{1}{G[0]}\int d^{d+1}x\,d^{d+1}y\,\bar{f}(\theta x)f(y)\frac{\delta}{\delta J(x)}\frac{\delta}{\delta J(y)}G[J]\right|_{J=0}\\
 & =\int d^{d+1}x\,d^{d+1}y\,\bar{f}(\theta x)S(x-y)f(y)\\
 & =\int d^{d+1}x\,d^{d+1}y\,\bar{f}(x)S(\theta x-y)f(y)
\end{align*}
where $S(x-y)$ denotes the Green function of the operator in the
quadratic from appearing in $S[\varphi]$ and the bar indicates complex
conjugation. Thus for this case, the requirement of reflection positivity
is equivalent to,
\begin{equation}
(\theta f,f)\geq0\;\;,Support(f)\subset\mathbb{R}_{d+1}^{+}\label{eq:rpf}
\end{equation}
where the scalar product $(,)$ is defined by,
\begin{equation}
(f,g)=\int d^{d+1}x\,d^{d+1}y\,\bar{f}(x)S(x-y)g(y)\label{eq:scprod}
\end{equation}
and $\theta f=f(\theta x)\;\;,\theta x=(-x_{d+1},\bar{x})$. It is
noted that this scalar product is defined not only for sources $f(x)$
with support on $\mathbb{R}_{d+1}^{+}$ but also for sources with
support in $\mathbb{R}_{d+1}$. The last two equations define what
is sometimes referred to as the reflection positivity requirement
for a one particle system\cite{guerra2006euclidean} or the RP requirement
for a bilinear form acting on a inner product space\cite{anderson2013defining}.
Motivated by the previous construction the external sources $f(x)$
are taken to belong to the Euclidean Hilbert space $\mathcal{E}$
of those tempered distributions\footnote{This allows to consider point-like external sources. }
in $d+1$-dimensional configuration space $\mathbb{R}_{d+1}$ , which
are square integrable with respect to the scalar product (\ref{eq:scprod}).
\begin{example}
\label{exa:1-sec-1}For the case of the free scalar field, 
\begin{align*}
S(x-y) & =\frac{1}{(2\pi)^{d+1}}\int d\bar{k}\,dk_{d+1}\frac{e^{i\bar{k}\cdot(\bar{x}-\bar{y})+ik_{d+1}(x_{d+1}-y_{d+1})}}{k_{d+1}^{2}+\bar{k}^{2}+m^{2}}
\end{align*}
This last expression shows that the poles in the $k_{d+1}$ complex
plane of $S(p)$, the propagator in momentum space, are located over
the imaginary axis at $\pm i\,\omega_{m}(\bar{k})$($\omega_{m}(\bar{k})=\sqrt{\bar{k}^{2}+m^{2}}$)
in the $k_{d+1}$ complex plane. The quantity $\omega_{m}(\bar{k})$
correspond to the energy eigenvalues for this one particle system
as will be shown below. 

Let us consider the case when $f(x)=\sum_{j}\,q_{j}\delta(x-r(j))\;,q_{i}\in\mathbb{\mathbb{C}}$
where $r(j)$ are the fixed positions of the \textquotedbl{}charge
distribution\textquotedbl{}\cite{Uhlmann1979}. Using (\ref{eq:scprod})
leads to,
\begin{align}
(\theta f,f) & =\intop d^{d+1}x\,d^{d+1}y\,\sum_{i,j}q_{i}^{*}q_{j}\delta(\theta x-r(i))S(x-y)\delta(y-r(j))\nonumber \\
 & =\sum_{i,j}q_{i}^{*}q_{j}S(\theta r(i)-r(j))\nonumber \\
 & =\int\frac{d\bar{k}}{(2\pi)^{d}}\sum_{i,j}a_{i}^{*}a_{j}\int\,\frac{dk_{d+1}}{2\pi}\frac{e^{-ik_{d+1}(r_{d+1}(i)+r_{d+1}(j))}}{k_{d+1}^{2}+\bar{k}^{2}+m^{2}}\label{eq:thetaff}
\end{align}
where the $a_{i}$ are defined as follows, 
\[
a_{i}=q_{i}e^{i\bar{k}\cdot\bar{r}(i)}
\]
the integral in the last equality in (\ref{eq:thetaff}) can be done
using the residue theorem, noting that $r_{d+1}(i)+r_{d+1}(j)>0$,
a convenient contour is one that runs along the real axis and closes
through a semicircle at infinity below the real axis, this leads to,
\[
(\theta f,f)=\int\frac{d\bar{k}}{(2\pi)^{d}}\sum_{i,j}\bar{a}_{i}a_{j}\frac{e^{-\omega_{m}(\bar{k})(r_{d+1}(i)+r_{d+1}(j))}}{2\omega_{m}(\bar{k})}
\]
defining,
\[
b_{i}^{m}=a_{i}\frac{e^{-\omega(\bar{k})r_{d+1}(i)}}{\sqrt{2\omega(\bar{k})}}
\]
leads to,
\[
(\theta f,f)=\int\frac{d\bar{k}}{(2\pi)^{d}}\sum_{i,j}\bar{b}_{i}^{m}b_{j}^{m}=\int\frac{d\bar{k}}{(2\pi)^{d}}\sum_{i,j}\bar{b}_{i}^{m}T_{ij}b_{j}^{m}
\]
where the matrix $T_{ij}$ is a matrix with all entries equal to $1$.
This matrix is positive semi-definite with eigenvalues given by,
\[
Eig(T)=(N,0,\cdots,0)
\]
where $N$ is the dimension of the matrix.
\end{example}
%\textasciitilde{}
\begin{example}
For the case of the higher derivative scalar field, $S(p)$ is given
by,
\[
S_{hd}(p)=\frac{M^{2}-m^{2}}{(p^{2}+m^{2})(p^{2}+M^{2})}=\frac{1}{p^{2}+m^{2}}-\frac{1}{p^{2}+M^{2}}
\]
This last expression shows that the poles of propagator in momentum
space $S_{hd}(p)$ in the complex $p_{d+1}$ plane, are located over
the imaginary axis at $\pm i\,\omega_{m}(\bar{p})$ and at $\pm i\,\omega_{M}(\bar{p})$
($\omega_{M}(\bar{p})=\sqrt{\bar{p}^{2}+M^{2}}$) . The quantities
$\omega_{m}(\bar{p})$ and $\omega_{M}(\bar{p})$ correspond to the
energy eigenvalues for this system as will be shown below. It is noted
that in the limit $M\to\infty$, the previous example is obtained.

Naively repeating the calculation as in the previous example, for
$f(x)=\sum_{j}\,q_{j}\delta(x-r(j))$, one obtains,
\begin{align*}
(\theta f,f){}_{hd} & =\int\frac{d\bar{k}}{(2\pi)^{d}}\sum_{i,j}\bar{a}_{i}a_{j}\left(\frac{e^{-\omega_{m}(\bar{k})(r_{d+1}(i)+r_{d+1}(j))}}{2\omega_{m}(\bar{k})}-\frac{e^{-\omega_{M}(\bar{k})(r_{d+1}(i)+r_{d+1}(j))}}{2\omega_{M}(\bar{k})}\right)\\
 & =\int\frac{d\bar{k}}{(2\pi)^{d}}\left(\sum_{i,j}\bar{b}_{i}^{m}T_{ij}b_{j}^{m}-\sum_{i,j}\bar{b}_{i}^{M}T_{ij}b_{j}^{M}\right)
\end{align*}
, this expression is not positive definite. In this respect it is
worth noting that the poles in the complex $p_{d+1}$-plane of $S_{hd}(p),$
indicate that two different particles are described by this correlator.
If sources are included only for the particle with mass $m$ then
reflection positive holds, however the other pole corresponds to a
scalar free particle of mass $M$ that does not satisfy reflection
positivity, it is a \textquotedbl{}ghost\textquotedbl{} with negative
norm.
\end{example}
%\textasciitilde{}
\begin{example}
\label{exa:Free-scalar-field-nid}Free scalar field on non-integer
dimensional space\cite{Trinchero:2012zn}. In this case,
\begin{equation}
S_{nid}(p)=\frac{\left(p^{2}+M^{2}\right)^{2\alpha}}{p^{2}+m^{2}}
\;\;,\alpha\in\mathbb{R\,},\alpha\geq0\;,m^{2}=M^{2}\frac{\alpha n}{1-2\alpha}\label{eq:snid}
\end{equation}
where $M$ is a infrared regulator, and $n\geq2$ is a positive integer.
This expression shows that $S_{nid}(p)$ has poles in the complex
$p_{d+1}$ plane over the imaginary axis at $\pm i\,\omega_{m}(\bar{p})$($\omega_{m}(\bar{p})=\sqrt{\bar{p}^{2}+m^{2}}$).
It also has two branch cuts starting at $p_{d+1}=\pm i \omega_M (\bar{p})$ and ending at $\pm i \infty$, which 
correspond to  a branch cut in the $p^{2}$ complex plane
on the negative real axis for\footnote{It is worth noting that here $p$ denotes the Euclidean momentum.}
$p^{2}<-M^{2}$. This is similar to what happens in a interacting
local theory when considering corrections to the two-point function.
For example in a $\varphi^{3}$ theory the one-loop correction involving
particles of mass $\mu$ has a branch cut over the negative real axis
for $p^{2}<-4\mu^{2}$. This is a consequence of the optical theorem
because $-4\mu^{2}$ is the $p^{2}$ value required to produce a pair
of those particles. In section \ref{sec:New-reflection-positive}
this analogy will be used to generate non-local reflection positive
Euclidean field theories.

Under the same assumptions as in the previous cases it holds that,

\begin{align}
(\theta f,f){}_{nid} & =\intop d^{d+1}x\,d^{d+1}y\,\sum_{i,j}q_{i}q_{j}\delta(\theta x-r(i))S_{nid}(x-y)\delta(y-r(j))\nonumber \\
 & =\int\frac{d\bar{p}}{(2\pi)^{d+1}}\sum_{i,j}\bar{a}_{i}a_{j}\int\,dp_{d+1} \frac{e^{-ip_{d+1}(r_{d+1}(i)+r_{d+1}(j))}\left(p_{d+1}^{2}+\bar{p}^{2}+M^{2}\right)^{2\alpha}}{p_{d+1}^{2}+\bar{p}^{2}+m^{2}}\nonumber \\
 &=\int\frac{d\bar{p}}{(2\pi)^{d+1}}\sum_{i,j}\bar{a}_{i}a_{j} \; I(r)
 \label{eq:offf}
\end{align}
where $r=r_{d+1}(i)+r_{d+1}(j)>0$ and ,
\[
a_{i}=q_{i}e^{i\bar{p}\cdot\bar{r}(i)}
\]
as before. The integral $I(r)$ over $p_{d+1}$ can be done by choosing a contour in
the complex $p_{d+1}$ plane, that runs bellow the real axis and continues
through a quarter of a circle at infinity on the lower half plane up to the negative imaginary axis, goes back and forth along this axis up to
$-i \omega_M (\bar{p})$ and returns to the real axis  through the other quarter of a circle at infinity on the lower half plane. The contribution
of both quarters of circle vanishes( because $r=r_{d+1}(i)+r_{d+1}(j)>0$ !). Circulating the contour in the clockwise sense,
the contribution of the back and forth path along the negative imaginary axis gives,
\begin{equation}
I_{cut}(r)= \int_{\omega_M ({\bar p})}^{\infty} d\rho\,(-2 sin(2\pi \alpha)) e^{ -\rho r}\frac{(\rho^2 -\omega_M ({\bar p})^2 )^{2\alpha}}{(\rho^2 -\omega_m ({\bar p})^2 )}
\end{equation}
which shows that for $0<\alpha < 1/2$  the integrand in this expression is negative for any value of $\rho$ within the integration limits. The  residue of the pole at $p_{d+1}=-i\omega_m ({\bar p})$ is,
\begin{eqnarray}
 && Res  \left[ \frac{(p_{d+1}^2 +\omega_M ({\bar p})^2 )^{2\alpha}}{(p_{d+1}^2 + \omega_m ({\bar p})^2 )} e^{-ip_{d+1} r}\right]_{p_{d+1}=-i\omega_m ({\bar p})}
=\nonumber\\
 && = i \frac{(M^2 - m^2 )^{2\alpha}}{2\omega_m ({\bar p})} e^{-\omega_m ({\bar p})r}
\end{eqnarray}
leading to,
\begin{eqnarray}
I(r) &=& -2\pi i Res  \left[ \frac{(p_{d+1}^2 +\omega_M ({\bar p})^2 )^{2\alpha}}{(p_{d+1}^2 + \omega_m ({\bar p})^2 )} e^{-ip_{d+1}  r}\right]_{p_{d+1}=-i\omega_m ({\bar p})} -I_{cut} \nonumber\\
&=& 2 \pi \frac{(M^2 - m^2 )^{2\alpha}}{2\omega_m ({\bar p})} e^{-\omega_m ({\bar p}) r} -I_{cut}
\label{I}\end{eqnarray}
which shows that $I >0$ for $M^2 - m^2>0 $. Defining,
\[
b_{i}=a_{i}\left(\frac{(M^{2}-m^{2})^{2\alpha}}{2\sqrt{\bar{p}^{2}+m^{2}}}\right)^{\frac{1}{2}}e^{-\omega_{m}(\bar{p})\,r_{d+1}(i)}\;\;,
b_i (\rho)= a_i e^{-\rho r_{d+1} (i)} \sqrt{\frac{(\rho^2 -\omega_M ({\bar p})^2 )^{2\alpha}}{(\rho^2 -\omega_m ({\bar p})^2 )}}
\]
leads to,
\[
(\theta f,f){}_{nid}=\int\frac{d\bar{p}}{(2\pi)^{d}} \left(\sum_{i,j} \bar{b}_{i}\,T_{ij}b_{j} +
\int_{\omega_M ({\bar p})}^{\infty} d\rho\, \sum_{i,j} \bar{b}_{i}(\rho)T_{ij}b_{j}(\rho) \right)
\]
using the same argument as the one employed for the case of the free scalar field, 
the two terms  in the r.h.s. of the last equation are positive definite for $M^{2}-m^{2}>0$. Using (\ref{eq:snid}) 
this last condition implies that $\alpha<\frac{1}{n+2}$, and therefore for these $\alpha$ values reflection 
positivity holds for the free scalar field in a NIDS.

Below an alternative proof of reflection positivity for the scalar
field on a non-integer dimensional space is presented. Using Newtons binomial
series the integral $I$ appearing in (\ref{eq:offf}) can be written as follows,
\begin{equation}
I(r)=\int_{-\infty}^{\infty} dp_{d+1} e^{-ip_{d+1}r} \sum_{j=0}^{\infty} \left(\begin{array}{c}
2\alpha\\
j
\end{array}\right)(p_{d+1}^2 +\omega_m)^{2\alpha-j-1} (\omega_M ({\bar p})^2 -\omega_m ({\bar p})^2 )^j
\end{equation}
where  $$\left(\begin{array}{c}
2\alpha\\
j
\end{array}\right)= \frac{2\alpha (2\alpha-1) \cdots ((2\alpha-j+1))}{j!}$$ 
and $r=r_{d+1}(i)+r_{d+1}>0$. The binomial series inside the integral is uniformly convergent for $p_{d+1}^2 >M^2 -2 m^2$,
therefore  by analytic continuation it coincides and converges uniformly to  $(p_{d+1}^2 + \omega_{M}^2)^{2\alpha}$ in 
the whole analyticity domain of this function, which includes the whole real axis. Thus integration and summation can be 
interchanged. Using that,
\begin{equation}
\int_{-\infty}^{\infty} dp_{d+1} e^{-ip_{d+1} r} (p_{d+1}^2 +\omega_m)^{2\alpha-j-1} =\frac{2^{2\alpha-j+
\frac{1}{2}}\sqrt{\pi}}{\Gamma(j+1-2\alpha)} K_{2\alpha-j-\frac{1}{2}}(r \omega_m)
\left( \frac{r}{\omega_m}\right)^{j-2\alpha +\frac{1}{2}}  
\end{equation}
leads to,
\begin{equation}\label{II}
I(r)=\sum_{j=0}^{\infty} \left(\begin{array}{c}
2\alpha\\
j
\end{array}\right)(M^2 -m^2 )^j \, \frac{2^{2\alpha-j+1/2}\sqrt{\pi}}{\Gamma(j+1-2\alpha)}  K_{2\alpha-j-1/2}(r \omega_m)
\left(\frac{r}{\omega_m}\right)^{j-2\alpha +1/2} 
\end{equation}
where $K_{\nu} (r) $ denotes the  modified  Bessel function such that $\lim_{r\to \infty}K_{\nu}(r)=0$.
For $M^2 - m^2 >0$ each term in this summation is positive thus implying that $I(r)>0$ in this case, 
using the same argument appearing after (\ref{I})  in the previous  proof, shows that reflection positivity 
holds for $M^2 - m^2 >0$, i.e. for $\alpha<\frac{1}{n+2}$.

For $\alpha<0$, which implies $d<n$, RP does not hold.
This result may be shown using the same methods as above. However, it
is important to note that for $\alpha<0$, the Fourier transform appearing in 
(\ref{eq:snid}) presents additional poles. In order to apply the same
calculation as in the $\alpha >0$ case, it is convenient to rewrite the Fourier transform
as follows,
\begin{equation}
S_{nid}(p)=\frac{\left(p^{2}+M^{2}\right)^{1+2\alpha}}{(p^{2}+m^{2})(p^{2}+M^{2})}\;\;,\alpha\in\mathbb{R\,},\alpha<0\;,m^{2}=M^{2}\frac{\alpha n}{1-2\alpha}
\end{equation} 
where now $1+2\alpha$, the exponent in the numerator, is positive. Recalling the previous example, 
this last expression shows that negative norm additional poles
 in the $p_{d+1}$-plane appear at  $\pm i\,\omega_{M}(\bar{p})$($\omega_{M}(\bar{p})=\sqrt{\bar{p}^{2}+M^{2}}$).
These poles spoil RP for $\alpha<0$. 
\end{example}  

\section{Scalar product and Energy\label{sec:Scalar-product-and}}

The pre-Hilbert\footnote{This is not a Hilbert because the inner product defined above is positive
semi-definite. To get a Hilbert space a quotient by the zero norm
states should be done.} 1-particle space $\mathcal{H}$, for a theory defined in Minkowski
space is the one of tempered distributions over $\mathbb{R}_{d+1}$
with support on a fixed $x_{d+1}=x_{d+1}^{0}$ slice, i.e. of the
form,
\[
f(x)=f(\bar{x})\delta(x_{d+1}-x_{d+1}^{0})
\]
equipped with the scalar product,
\[
<g,f>=(\theta g,f)\;\;,
\]
this scalar product is positive semi-definite by virtue of RP. Taking
$x_{d+1}^{0}=0$, gives,
\begin{equation}
<g,f>=\int\,d\bar{x}\,d\bar{y}\,g^{*}(\bar{x})S((0,\bar{x})-(0,\bar{y}))f(\bar{y})\label{eq:scp}
\end{equation}
which in terms of the d-dimensional Fourier transforms of $g^{*}(\bar{x}),\;f(\bar{y})$ and $S((0,\bar{x})-(0,\bar{y}))$
is given by,
\begin{equation}\label{sc}
<g,f>=\int\, \frac{d\bar{k}}{(2\pi)^d}\,g^*({\bar k}) S({\bar k}) f({\bar k})
\end{equation}

\noindent The energy eigenvalues can be obtained using the scalar product by means of,
\begin{eqnarray}
<g,Hf> &=& \left.-\frac{d}{d\tau}(g,\theta T_{\tau}f)\right|_{\tau\to 0^+}\nonumber\\
&=&\int\,d\bar{x}\,d\bar{y}\,g^{*}(\bar{x})\left[-\frac{d}{d\tau}S((0,\bar{x})-(\tau,\bar{y}))\right]_{\tau\to 0^+} f(\bar{y})\label{eq:scp}
\end{eqnarray}
where $T_{\tau}$ is the traslation by $\tau$ in the $d+1$ direction\footnote{Althought this construction 
is aimed at defining the theory in Minkowki space, for the sake of simplicity, 
the Euclidean notation will still be employed in what follows.},
i.e.,
\[
T_{\tau}f(x_{d+1},\bar{x})=f(x_{d+1}-\tau,\bar{x})
\]
which in terms of Fourier transforms is given by,
\begin{equation}\label{h}
<g,H f>=\int\, \frac{d\bar{k}}{(2\pi)^d}\,g^*({\bar k}) D_{\tau}S({\bar k},\tau)|_{\tau \to 0^+} f({\bar k})
\end{equation}
where $D_{\tau}S({\bar k},\tau)$ is defined by,
\begin{equation}
-\frac{d}{d\tau}S((0,\bar{x})-(\tau,\bar{y}))= \int\, \frac{d\bar{k}}{(2\pi)^d} D_{\tau}S({\bar k},\tau)\,
e^{-i {\bar k}\cdot (\bar{x}-\bar{y})}
\end{equation}
The energy eigenvalues can be obtained using (\ref{sc}) and (\ref{h}) for momentum eigenstates 
$f_{\bar{K}}(\bar{k})=\delta(\bar{k}-\bar{K})$ as follows,
\begin{equation}\label{E}
 E({\bar{K}})=\frac{<g,Hf_{\bar{K}} >}{<g,f_{\bar{K}} >}
\end{equation}

\begin{example}
\emph{The free scalar field}. Taking $x_{d+1}^{0}=0$ and noting that,
\begin{eqnarray}
S((0,\bar{x})-(\tau,\bar{y})) & =&\frac{1}{(2\pi)^{d+1}}\int d\bar{k}\,dk_{d+1}\frac{e^{-i\bar{k}\cdot(\bar{x}-\bar{y})+ik_{d+1}\tau}}{k_{d+1}^{2}
+\bar{k}^{2}+m^{2}}\nonumber\\
 & =&\int\frac{d\bar{k}}{(2\pi)^{d}}\,\frac{e^{-i\bar{k}\cdot(\bar{x}-\bar{y})}e^{-\omega_{m}(\bar{k})\tau}}{2\omega_{m}(\bar{k})}\;\;,
 \;\omega_{m}(\bar{k})=\sqrt{\bar{k}^{2}+m^{2}}
\label{2fs}\end{eqnarray}
 leads in this case to the positive definite scalar product,
\[
<g,f>=\int\frac{d\bar{k}}{(2\pi)^{d}}\,\frac{g(\bar{k})^{*}f(\bar{k})}{2\omega_{m}(\bar{k})}
\]

The one-particle energy eigenvalues can be computed using (\ref{E}), in order to do so it is noted that, 
\begin{align*}
(g,\theta T_{\tau}f) & =\int d\bar{x}\,d^{d+1}y\,g^{*}(\bar{x})S((0,\bar{x})-y)f(\bar{y})\,\delta(y_{d+1}-\tau)\\
 & =\int\,d\bar{x}\,d\bar{y}\,g^{*}(\bar{x})\,S((0,\bar{x})-(\tau,y)\,f(\bar{y})\\
 & =\int\frac{d\bar{k}}{(2\pi)^{d}}\,\frac{g(\bar{k})^{*}f(\bar{k})}{2\omega_{m}(\bar{k})} e^{-\omega_{m}(\bar{k})\tau}
\end{align*}
for the case of a plane wave $f(\bar{k})=\delta(\bar{k}-\bar{K})$,
this leads to,
\[
\frac{<g,Hf>}{<g,f>}=\frac{\int\frac{d\bar{k}}{(2\pi)^{d}}\,g(\bar{k})^{*}f(\bar{k})}{\int\frac{d\bar{k}}{(2\pi)^{d}}\,\frac{g(\bar{k})^{*}f(\bar{k})}{2\omega_{m}(\bar{k})}}=\omega_{m}(\bar{K})
\]

\end{example}
\textasciitilde{}
\begin{example}
\emph{Higher derivative scalar field.} Separating the propagator in
the two parts,
\[
S_{hd}(p)=S_{hd}^{(m)}(p)+S_{hd}^{(M)}(p)\;\;,S_{hd}^{(m)}(p)=\frac{1}{p^{2}+m^{2}}\;,S_{hd}^{(M)}(p)=-\frac{1}{p^{2}+M^{2}}
\]
then the calculation for each pole is the same as in the previous
case, leading to the following scalar products and energies,
\begin{align*}
 & <g,f>_{m}=\int\frac{d\bar{k}}{(2\pi)^{d}}\,\frac{g(\bar{k})^{*}f(\bar{k})}{2\omega_{m}(\bar{k})}\;\;,E_{m}=\omega_{m}(\bar{k})\\
 & <g,f>_{M}=-\int\frac{d\bar{k}}{(2\pi)^{d}}\,\frac{g(\bar{k})^{*}f(\bar{k})}{2\omega_{m}(\bar{k})}\;\;,E_{M}=\omega_{M}(\bar{k})
\end{align*}
which shows that $S_{hd}^{(M)}(p)$ leads to a negative scalar product
in accordance with the fact that for this part reflection positivity
does not hold. 

\noindent It is important to note that if this separation of the contribution
for each pole is not done then the result for the energy eigenvalues
is wrong. Indeed using the whole propagator one gets,
\[
(g,\theta T_{\tau}f)=\int\frac{d\bar{k}}{(2\pi)^{d}}\,g(\bar{k})^{*}f(\bar{k})\left(\frac{e^{-\omega_{m}(\bar{k})\tau}}{2\omega_{m}(\bar{k})}-\frac{e^{-\omega_{M}(\bar{k})\tau}}{2\omega_{M}(\bar{k})}\right)
\]
which implies,
\[
\left.-\frac{d}{d\tau}(g,\theta T_{\tau}f)\right|_{\tau \to 0^+}=0\Rightarrow H=0
\]
\end{example}

\textasciitilde{}
\begin{example}
\emph{Free scalar field on non-integer dimensional space}. Using the
results in example \ref{exa:Free-scalar-field-nid} leads to,
\[
S_{nid}((0,\bar{x})-(\tau,\bar{y}))=\int\frac{d\bar{p}}{(2\pi)^{d+1}}\; I(\tau)
\]
The expression for $I(\tau)$ given in (\ref{I}) will be employed in this calculation, i.e.,
\begin{eqnarray}
 I(\tau)&=& \pi \frac{(M^2 - m^2 )^{2\alpha}}{\omega_m ({\bar p})} e^{-\omega_m ({\bar p}) |\tau|} +
 2  \sin(2\pi \alpha) \nonumber \\
&& \times \int_{\omega_M ({\bar p})}^{\infty} d\rho\, e^{ -\rho \tau}
\frac{(\rho^2 -\omega_M ({\bar p})^2 )^{2\alpha}}{(\rho^2 -\omega_m ({\bar p})^2 )}
\end{eqnarray}
The integral appearing in the r.h.s. of the 
last equation can be explicitely computed by expanding the 
integrand as follows,
\begin{eqnarray}
 I(\tau)&=& \pi \frac{(M^2 - m^2 )^{2\alpha}}{\omega_m ({\bar p})} e^{-\omega_m ({\bar p}) |\tau|}+ 2 \sin(2\pi \alpha)\nonumber\\
 && 
 \times \int^{\infty}_{\omega_M}\,d\rho \sum^{\infty}_{k=0} \frac{(-\rho\tau)^k}{k!}\sum^\infty_{j=0}
\left(\begin{array}{c}
2\alpha\\
j
\end{array}\right)
\frac{(-\omega^2_M)^j \rho^{2(2\alpha-j)}}{\rho^2 - \omega^2_m}  
 \end{eqnarray}
the two infinite series appearing inside the integral in the r.h.s. of the last equation are uniformly convergent,
the first one converges to $e^{-\rho \tau}$  and the other is uniformly convergent for $\rho^2>\omega^2_M$ which is 
the integration interval. Therefore the order of integration and summation can be interchanged. The resulting integrals 
can be computed leading to,
\begin{equation}
 I(\tau)= \pi \frac{(M^2 - m^2 )^{2\alpha}}{\omega_m ({\bar p})}e^{-\omega_m ({\bar p})|\tau|}+
 \omega_m({\bar p})^{4\alpha -1}\sum^{\infty}_{k=0}
 \frac{(-\tau \omega_m({\bar p}))^k}{k!}
 I(k,{\bar p}) 
 \end{equation}
where the adimensional quantity $I(k,{\bar p})$ is given by,
\begin{eqnarray}
 I(k,{\bar p})&=& \sin(2\pi \alpha) 
  \sum^\infty_{j=0}
\left(\begin{array}{c}
2\alpha\\
j
\end{array}\right)
\left(-\frac{\omega_M({\bar p})}{\omega_m ({\bar p})}\right)^{2j}  \nonumber\\
&& \times  Beta\left(\frac{\omega_m({\bar p})^2}{\omega_M({\bar p})^2} , -\frac{(4\alpha+k-2j-1)}{2},0\right)
\end{eqnarray}
where  $Beta$ denotes Euler's beta function. Using these results in  eqs. (\ref{sc}), (\ref{h}) and  replacing in (\ref{E})
leads to,
\begin{equation}
 E(\bar{K})=\omega_m(\bar{K}) \left(
 \frac{\pi(M^2 -m^2)^{2\alpha}\omega_m(\bar{K})^{-4\alpha}+ I(1,\bar{K})}
{\pi(M^2 -m^2)^{2\alpha}\omega_m(\bar{K})^{-4\alpha}
 +I(0,{\bar K})}\right)
\end{equation}
Expanding the quantity multiplying $\omega_m(\bar{K})$ in the r.h.s. of this last expression in powers 
of $\alpha$ leads to,
\begin{equation}
E(\bar{K})=\omega_m(\bar{K})\left[1-2 \alpha\,Beta\left( \frac{\bar{K}^2}{\bar{K}^2+ M^2},\frac{1}{2},0 \right)
+{\mathcal{O}}(\alpha^2)\right] 
\end{equation}
to this order the rest mass is $m$, i.e. $E(0)=m$. However at higher orders there are 
non-vanishing corrections  to this last equation.
\end{example}

\section{Non-local reflection positive theories\label{sec:New-reflection-positive}}

Motivated by the similarity mentioned in example \ref{exa:Free-scalar-field-nid}
in section \ref{sec:Reflection-positivity} the following question
is addressed. Under what conditions the corrected Euclidean propagator
coming from a local interacting field theory leads to a scalar product,
defined as in (\ref{eq:scp}), satisfying the requirement of reflection
positivity? It is stressed that the interest is not in whether the
underlying local field theory satisfies reflection positivity or not,
but on whether a non-local field theory defined by a non-local action
obtained from the inverse of the corrected propagator satisfies reflection
positivity or not. In order to answer the question above, the corrected
Euclidean propagator is considered in greater detail. It is given
by,
\[
S_{c}(p)=\frac{1}{p^{2}+m^{2}-\Sigma(p^{2})}
\]
where in the perturbative case $\Sigma(p^{2})$ has an expansion in
powers of a coupling constant $g$ of the form,
\begin{equation}
\Sigma(p^{2})=g\Sigma_{1}(p^{2})+g^{2}\Sigma_{2}(p^{2})+\cdots.\label{eq:pert-sigma}
\end{equation}
 The propagator in coordinate space is given by,
\[
S_{c}(x-y)=\frac{1}{(2\pi)^{d+1}}\int d\bar{p}\,dp_{d+1}\frac{e^{i\bar{p}\cdot(\bar{x}-\bar{y})+ip_{d+1}(x_{4}-y_{4})}}{p_{d+1}^{2}+\bar{p}^{2}+m^{2}-\Sigma(p^{2})}
\]
It is useful to consider the expansion of $\Sigma(p^{2})$ around
$p^{2}=-m^{2}$, which leads to the following identity,
\[
p^{2}+m^{2}-\Sigma(p^{2})=p^{2}+m^{2}-\Sigma(-m^{2})-\Sigma'(-m^{2})(p^{2}+m^{2})-\frac{\Sigma''(-m^{2})}{2!}(p^{2}+m^{2})^{2}-\cdots
\]
the equation, 
\[
p^{2}+m^{2}-\Sigma(p^{2})=0
\]
enables to compute the poles of $S_{c}(p)$. This equation together
with (\ref{eq:pert-sigma}) shows that the location of these poles
is such that $p^{2}+m^{2}\simeq\mathcal{O}(g)$. Thus up to $\mathcal{O}(g^{2})$
the poles in the $p_{d+1}$ complex plane can be obtained solving
the following equation, whose solutions are also shown below,
\[
p^{2}+m^{2}-\Sigma(-m^{2})=0\;\;\Rightarrow p_{d+1}=
\pm i\left(\omega(\bar{p})-\frac{\Sigma(-m^{2})}{2\omega(\bar{p})}\right)\;.
\]
Evaluating the integral over $p_{d+1}$ using residues leads to,
\[
S_{c}(x-\theta y)=\frac{i}{(2\pi)^{d}}\int d\bar{p}e^{-i\bar{p}\cdot(\bar{x}-\bar{y})}
\frac{e^{-(\omega(\bar{p})-\frac{\Sigma(-m^{2})}{2\omega(\bar{p})})(x_{4}-\theta y_{4})}}{2i\left(\omega(\bar{p})
-\frac{\Sigma(-m^{2})}{2\omega(\bar{p})}\right)}\;\;,x_{4}>0,y_{4}<0
\]
using the same arguments as in the previous cases it is concluded
that RP will hold whenever the quantity $\omega(\bar{p})-\frac{\Sigma(-m^{2})}{2\omega(\bar{p})}$
is real and greater than zero. If $\Sigma(-m^{2})$ is real then this
will be the case because in the perturbative regime $\Sigma(-m^{2})\ll\omega(\bar{p})^{2}$.
However if $\Sigma(-m^{2})$ has a imaginary part then RP would not
hold. As is well known, this can happen even if the underlying local
field theory satisfies RP. Indeed in that case such a imaginary part
will appear if the particle whose corrected propagator is considered
can decay through a real process into other particles\cite{Veltman:1963th},
i.e. if it is a unstable particle. This is a consequence of the optical
theorem for the underlying local RP theory. In addition, there exists
a reduction to the absurd argument leading to the same result. Suppose
that RP would hold for the corrected propagator, that would imply
that the time evolution for this corrected particle is unitary, however
this is not true since the $S$-matrix is not unitary for a theory
including unstable asymptotic particles\cite{Veltman:1963th}. Furthermore,
this makes sense from the physical point of view, if the particle
can decay then its energy contains an imaginary part and the time
evolution restricted to this one \emph{unstable} particle subspace
is not unitary. 

It is possible to go a step further in the analogy between the corrected
propagator and the free theory in a NIDS. The two point function for
the later case can be written as follows,
\begin{align*}
S_{nid}(p) & =\frac{(p^{2}+M^{2})^{2\alpha}}{p^{2}+m^{2}}=\frac{1}{(p^{2}+m^{2})e^{-2\alpha\log(p^{2}+M^{2})}}\\
 & =\frac{1}{p^{2}+m^{2}-\Sigma^{(nid)}(p^{2})}
\end{align*}
where,
\begin{equation}
\label{alfa}
\Sigma^{(nid)}(p^{2})=\alpha\Sigma_{1}^{(nid)}(p^{2})+\alpha^{2}\Sigma_{2}^{(nid)}(p^{2})+\cdots
\end{equation}
with,
\begin{align*}
\Sigma_{1}^{(nid)}(p^{2}) & =-2(p^{2}+m^{2})\log(p^{2}+M^{2})\;\;,\Sigma_{2}^{(nid)}(p^{2})=2 (p^{2}+m^{2})\log(p^{2}+M^{2})^{2}
\end{align*}
comparing (\ref{eq:pert-sigma}) with (\ref{alfa}) shows that the departure $\alpha$ from
the integer dimensional case plays the role of a coupling constant
in this analogy. %
\begin{comment}
It may happen that the two point function for the free theory on a
NIDS can be obtained as the corrected two point function of a interacting
local field theory on a integer dimensional space, however this is
for the moment an open question.
\end{comment}

\begin{comment}
The propagator for the NID case is given by,
\begin{align*}
S_{nid}(p) & =-i\frac{(p^{2}+M^{2})^{\alpha}}{p^{2}+m^{2}}=\frac{-i}{(p^{2}+m^{2})e^{-\alpha\log(p^{2}+M^{2})}}\\
 & =\frac{-i}{p^{2}+m^{2}-\Sigma^{(nid)}(p^{2})}
\end{align*}
where,
\begin{equation}
\label{alfa}
\Sigma^{(nid)}(p^{2})=\alpha\Sigma_{1}^{(nid)}(p^{2})+\alpha^{2}\Sigma_{2}^{(nid)}(p^{2})+\cdots
\end{equation}
with,
\begin{equation}
\Sigma_{1}^{(nid)}(p^{2}) & =-(p^{2}+m^{2})\log(p^{2}+M^{2})\;\;,\Sigma_{2}^{(nid)}(p^{2})=\frac{1}{2}(p^{2}+m^{2})\log(p^{2}+M^{2})^{2}
\end{equation}
comparing (\ref{eq:pert-sigma}) with (\ref{alfa}) shows that the departure $\alpha$ from
the integer dimensional case plays the role of a coupling constant
in this analogy. 
\end{comment}

\section{Concluding remarks\label{sec:Concluding-remarks}}

In this work the requirement of RP has been studied in some particular
cases. This study leads to the following conclusions,
\begin{itemize}
\item The pole structure of the corresponding two point functions encodes
the states present in the theory. In particular, the case of the higher
derivative scalar field is a clear illustration that in order to understand
the spectrum it is convenient (when possible) to separate the Fourier
transformed two point function in single pole terms.
\item The scalar theory defined on a NIDS satisfies RP for $0<\alpha<\frac{1}{n+2}$, i.e. for $n<d$
. This result is extremely
interesting because it implies that a field theory on a NIDS produces
a unitary evolution. Having a unitary evolution for a field theory
on a NIDS space permits to address the question of whether physical
space has integer dimension or not.

On the other hand for $\alpha<0$, i.e. for $d<n$ RP does not hold. 
This result agrees partially with the one in \cite{Rychkov-16},
however in this respect it is important to realize that the NIDS
employed in both works are not the same.
  
\item The appearance of a branch cut for the two point function in momentum
space for the NIDS has suggested an analogy with an interacting perturbative
local field theory. In this analogy, the deformation parameter\footnote{This quantity parametrizes the deviation from the integer dimensional
case, as shown by equation (\ref{eq:da}).} $\alpha$ in the NIDS plays the role the coupling constant
in the interacting theory. The question of whether there exists interacting
perturbative local field theories leading to the same two point function
as the free theory on the NIDS remains open. It has been shown that
the corrected propagator of a local unitary field theory leads to
a reflection positive Euclidean non-local theory whenever the particle
whose corrected propagator is considered can not decay through a real
process into other particles.
\end{itemize}
\textbf{Acknowledgments. }

I would like to thank H. Casini for explaining to me the relation between
RP and unitarity, and also for a useful critical reading of the manuscript. 
I also thank G. Torroba for encouraging discussions on the subject and  
to S. Grillo for his mathematical support. I appreciate
discussions with S. Rychkov.
\section*{Appendix. }

\subsection*{Spectral functions. }

The spectral function $\rho$ is related to the Fourier transform
of the two-point function $S(p^{2})$(a recent review of the subject
appears in \cite{Zwicky:2016lka}). It will be assumed that the singularities
of the Fourier transform of the Euclidean two-point function are given
by a pole at $p^{2}=-m^{2}$ and a cut starting at a certain real
negative value $-\mu^{2}$ of $p^{2}$. Using a keyhole contour $\gamma$
consisting of a circle at infinity and a back and forth path going
above and below the negative real axis starting at the first singularity
to the right, $S(p^{2})$ can be written as follows,
\[
S(p^{2})=\frac{1}{2\pi i}\int_{\gamma}ds\,\frac{S(s)}{s-p^{2}}
\]
assuming that $S(p^{2})$ is such that the circle at infinity does
not contribute to the integral in the last equation, then,
\[
S(p\text{\texttwosuperior})=\frac{-1}{2\pi i}\lim_{\epsilon\to0^+}\int_{-s_{0}}^{-\infty}ds\,\frac{S(s+i\epsilon)-S(s-i\epsilon)}{s-p^{2}}
\]
where $-s_{0}$ is a value of $s$ such that no singularities appear
for $s>-s_{0}$. The quantity multiplying $\frac{1}{s-p^{2}}$ is the
spectral function\footnote{For the case in which the two-point function is such that $S(s)$
is real for $s$ in a open interval of the real line, then Schwarz
reflection principle implies that,
\begin{equation}
\rho(s)=-\frac{1}{2\pi i}\left(\lim_{\epsilon\to0^+}S(s+i\epsilon)-S(s-i\epsilon)\right)=
-\frac{1}{\pi}\lim_{\epsilon\to0}Im[S(s+i\epsilon)]\label{eq:rho-im}
\end{equation}
}, i.e.,
\begin{equation}
\rho(s)=-\frac{1}{2\pi i}\left(\lim_{\epsilon\to0^+}S(s+i\epsilon)-S(s-i\epsilon)\right)\label{eq:spec-func}
\end{equation}
Reflection positivity in terms of the spectral function is the requirement
that,
\[
\rho(s)\geq0\;\;,\,s\in\mathbb{R}.
\]
Bellow the case of the scalar field in the NIDS is considered. In this case the circle at infinity does not contribute.
Using (\ref{eq:spec-func}) or (\ref{eq:rho-im}) leads to,
\begin{eqnarray}
\rho(s) & =&-\frac{1}{\pi}\lim_{\epsilon\to0^+}Im[S_{nid}(s+i\epsilon)]=
-\frac{1}{2\pi i}\left(\lim_{\epsilon\to0^+}S(s+i\epsilon)-S(s-i\epsilon)\right)\nonumber\\
 & =& \frac{1}{\pi}\lim_{\epsilon\to0^+}\frac{\epsilon((s+M^{2})^2+\epsilon^2)^{\alpha}}{(s+m^{2})^{2}+\epsilon^{2}}\nonumber\\
 &&\left( (s+m^2 + i \epsilon) e^{i 2\alpha \arctan(\frac{\epsilon}{s+M^2})}-
 (s+m^2 - i \epsilon) e^{-i 2\alpha \arctan(\frac{\epsilon}{s+M^2})}\right)\nonumber\\
 & =& \frac{1}{\pi}\lim_{\epsilon\to0^+}
 \frac{\epsilon((s+M^{2})^2+\epsilon^2)^{\alpha}}{(s+m^{2})^{2}+\epsilon^{2}}\times\nonumber\\
 & &\times \left(\epsilon \cos (2\alpha \arctan(\frac{\epsilon}{s+M^2}))-
 (s+m^2 )\sin(2\alpha \arctan(\frac{\epsilon}{s+M^2})\right)
\end{eqnarray}
which satisfies RP for $M^2>m^2$, since in that case it is  positive semi-definite for 
$s\in\mathbb{R}$, $\epsilon \ll 1$ and $\epsilon>0$.

\begin{comment}
\emph{Perturbatively corrected theory}. In this case $S(p^{2})$ is
given by,
\[
S_{c}(s)=\frac{1}{s+m^{2}-\Sigma(p^{2})}\underset{_{\mathcal{O}(g^{2})}}{=}\frac{1}{s+m^{2}-\Sigma(-m^{2})}
\]
where the last equality is valid only up to terms of order $g^{2}$
as explained in section \ref{sec:New-reflection-positive}. Next the
spectral function for this case is computed using\footnote{If $\Sigma(-m^{2})$ has a non-vanishing imaginary part then the property
in the previous footnote is not satisfied, and the equality (\ref{eq:rho-im})
does not hold.} (\ref{eq:spec-func}),
\begin{align*}
\rho_{c}(s) & =\frac{-1}{2\pi i}\left(\lim_{\epsilon\to0}S_{c}(s+i\epsilon)-S_{c}(s-i\epsilon)\right)\underset{_{\mathcal{O}(g^{2})}}{=}\frac{1}{2\pi i}\lim_{\epsilon\to0}\frac{2i\epsilon}{(s+m^{2}-\Sigma(-m^{2}))^{2}+\epsilon^{2}}\\
 & =\frac{1}{\pi}\lim_{\epsilon\to0}\left(\frac{\epsilon}{(s+m^{2})^{2}+\epsilon^{2}}+\frac{2(s+m^{2})\Sigma(-m^{2})\epsilon}{((s+m^{2})^{2}+\epsilon^{2})^{2}}+\mathcal{O}(\Sigma^{2})\right)
\end{align*}
which shows that if $Im[\Sigma(-m^{2})]\neq0,$ then the second term
gives a imaginary part to the spectral function which therefore violates
RP.
\end{comment}

\begin{comment}
\footnote{This follows from the fact that,
\[
M^{2}-m^{2}=M^{2}(1-\frac{\alpha n}{1-2\alpha})=M^{2}\left(\frac{1-\alpha(n+2)}{1-2\alpha}\right)
\]
is positive for $\alpha<\frac{1}{n+2}$ .}

\end{comment}

\bibliographystyle{unsrt}
\addcontentsline{toc}{section}{\refname}
\bibliography{Bibliography}

\end{document}